\documentclass[conference]{IEEEtran}
\IEEEoverridecommandlockouts
\usepackage{cite}
\usepackage{amsmath,amssymb,amsfonts}
\usepackage{algorithmic}
\usepackage{graphicx}
\usepackage{textcomp}
\usepackage{xcolor}
\usepackage{svg}
\usepackage{graphicx} 
\usepackage{subcaption} 
\def\BibTeX{{\rm B\kern-.05em{\sc i\kern-.025em b}\kern-.08em
    T\kern-.1667em\lower.7ex\hbox{E}\kern-.125emX}}
\begin{document}

\title{Dense-TSNet: Dense Connected Two-Stage Structure for Ultra-Lightweight Speech Enhancement\\
\thanks{Source codes will be released after paper acceptance, audio samples can be found at https://github.com/huaidanquede/Dense-TSNet}
}


\author{\IEEEauthorblockN{1\textsuperscript{st} Zizhen Lin}
\IEEEauthorblockA{\textit{School of Electronic Information} \\
\textit{Sichuan University}\\
Chengdu, China \\
linzizhen17@163.com}
\and

\IEEEauthorblockN{2\textsuperscript{nd} Yuanle Li}  
\IEEEauthorblockA{\textit{School of Communications and Information Engineering} \\  
\textit{Chongqing University of Posts and Telecommunications} \\  
Chongqing, China \\  
S220132086@stu.cqupt.edu.cn}  

\IEEEauthorblockN{3\textsuperscript{rd} Junyu Wang}
\IEEEauthorblockA{\textit{College of Intelligence and Computing} \\
\textit{Tianjin University}\\
Tianjin, China \\
junyu$\_$wang21@tju.edu.cn}
\and
\IEEEauthorblockN{4\textsuperscript{th} Ruili Li}
\IEEEauthorblockA{\textit{School of Medicine} \\
\textit{Tohoku University}\\
Sendai, Japan \\
li.ruili.t3@dc.tohoku.ac.jp}
}

\maketitle

\begin{abstract}
Speech enhancement aims to improve speech quality and intelligibility in noisy environments. Recent advancements have concentrated on deep neural networks, particularly employing the Two-Stage (TS) architecture to enhance feature extraction. However, the complexity and size of these models remain significant, which limits their applicability in resource-constrained scenarios. Designing models suitable for edge devices presents its own set of challenges. Narrow lightweight models often encounter performance bottlenecks due to uneven loss landscapes. Additionally, advanced operators such as Transformers or Mamba may lack the practical adaptability and efficiency that convolutional neural networks (CNNs) offer in real-world deployments. To address these challenges, we propose Dense-TSNet, an innovative ultra-lightweight speech enhancement network. Our approach employs a novel Dense Two-Stage (Dense-TS) architecture, which, compared to the classic Two-Stage architecture, ensures more robust refinement of the objective function in the later training stages. This leads to improved final performance, addressing the early convergence limitations of the baseline model.  We also introduce the Multi-View Gaze Block (MVGB), which enhances feature extraction by incorporating global, channel, and local perspectives through convolutional neural networks (CNNs). Furthermore, we discuss how the choice of loss function impacts perceptual quality. Dense-TSNet demonstrates promising performance with a compact model size of around 14K parameters, making it particularly well-suited for deployment in resource-constrained environments.
\end{abstract}

\begin{IEEEkeywords}
speech enhancement, lightweight, dense connection, multi-view, metric loss
\end{IEEEkeywords}

\section{Introduction}

Speech enhancement and noise suppression technologies aim to improve speech clarity in noisy environments \cite{pascual2017segan}. These technologies are vital for applications like voice calls, video recording, and automatic speech recognition (ASR). 

Recent research, inspired by TSTNN \cite{tstnn}, utilizes the Two-Stage (TS) architecture for deep feature extraction, which represents the Classic Two-Stage in this paper. This architecture reshapes and partitions spectral data, effectively addresses spectral distribution characteristics. TSTNN also features an encoder-decoder with a dilated dense block \cite{dilateddensenet}, which improves information flow stability and enhances local and large-scale feature dependency.

In the time-frequency (TF) domain, some models focus solely on amplitude information, while others enhance complex spectrograms (real and imaginary parts) to indirectly improve both amplitude and phase. For instance, CMGAN \cite{cmgan_interspeech} uses a dual-decoder structure for amplitude and complex spectrograms, and MP-SEnet \cite{mpsenet_interspeech} directly estimates amplitude and phase spectrograms.

Loss functions in speech enhancement typically compute the \(L_{p}\)-norm distance between estimated and target spectrograms. MetricGAN \cite{metricgan,metricgan+}, however, uses learned evaluation metrics from a discriminator to address the challenge of non-differentiable assessment metrics.

Despite the strong performance of deep neural network-based methods, these come with increased model complexity. New methods like Transformer \cite{vaswani2017attention} or Mamba \cite{mamba,se-mamba} also face challenges with optimized deployment operators. Consequently, developing ultra-lightweight models is essential for practical applications in edge devices such as smartphones and cameras. Acknowledging the current landscape, research on single-channel speech enhancement models with around 10k parameters is still limited. This gap underscores the need for innovative approaches to develop effective models at this scale.

Motivated by this, we propose an ultra-lightweight speech enhancement network called Dense-TSNet. Our contributions are as follows:

\begin{figure*}[t]
  \centering
  \vspace{-0.4cm}
  \includegraphics[width=\linewidth]{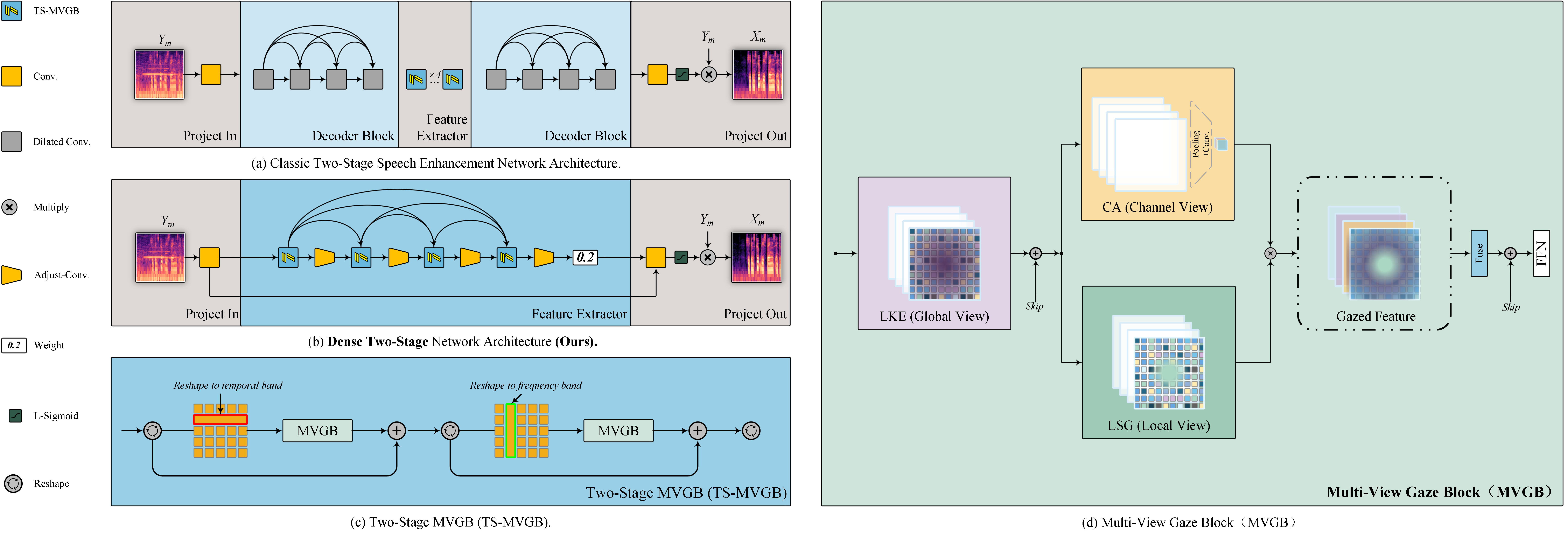}
  \caption{The overall architecture of Classic Two-Stage and \textbf{Dense-TS (ours)}.}
  \label{fig:architecture}
  \vspace{-0.4cm}
\end{figure*} 

\begin{itemize}
    \item We propose a novel Dense-TS (Dense Two-Stage) structure to address the issue of loss landscape \cite{loss_landscape} smoothness in narrow lightweight models.

    \item We design the Multi-View Gaze Block (MVGB) to capture features from global, channel, and local views, entirely based on CNNs \cite{CNNgradient} for ease of deployment.
   
    \item We explore loss functions through comparative experiments and propose a method utilizing Consistency Magnitude Loss to reconstruct speech signals, as relying on Metric Loss to improve objective perceptual metrics may negatively impact subjective perception. 
    \item The proposed Dense-TS Net model has a size of only 14K, which, to our knowledge, is the first to achieve a speech enhancement model with approximately 10k parameters.

\end{itemize}

\section{methods}
\subsection{Overall Structure}
The speech signal is processed using Short-Time Fourier Transform (STFT) to convert it into the time-frequency (TF) domain. The magnitude and phase spectra are derived from the complex spectrum. As illustrated in Fig. \ref{fig:architecture}, Dense-TSNet inputs only the magnitude spectrum and performs time-frequency feature extraction through the Dense-TS structure with the TS-MVGB.

\subsection{Dense-TS}

The classic Two-Stage speech enhancement network architecture \cite{tstnn}, shown in Fig. \ref{fig:architecture} (a), consists of two core components. The first is a densely connected encoder-decoder with dilated convolutions and dense connections for robust feature fusion. The second is a feature extraction layer with serially connected Two-Stage modules as shown in Fig. \ref{fig:architecture} (c), decomposing spectral feature extraction into two phases. Given a spectral feature map \( D \in \mathbb{R}^{B \times T \times F \times C} \), \( D \) is reshaped to \( D_T \in \mathbb{R}^{BF \times T \times C} \) for temporal dependencies and then reshaped to \( D_F \in \mathbb{R}^{BT \times F \times C} \) for frequency dependencies, with the final output \( D_o \) reshaped to the input size. This architecture performs well with a channel count of 64 \cite{wang23DPCFCS_interspeech,cmgan_interspeech,mpsenet_interspeech,DB-AIATinterspeech12} but faces limitations for ultra-lightweight networks with a narrow architecture with fewer channels, which may cause uneven loss landscapes \cite{loss_landscape}.

To address these issues, we propose the Dense-TS structure, shown in Fig. \ref{fig:architecture}(b). This structure replaces dilated convolutions in Dilated Dense Block with Two-Stage modules, resulting in a network composed entirely of densely connected Two-Stage modules. 

Specifically, we mapping features to a high-dimensional space using 1×1 convolutions. Unlike the classic Two-Stage structure, we omit the dilated dense block in the encoder and directly proceed to the Two-Stage feature extraction block with dense connection to merge shallow and deep features. The Two-Stage structure remains consistent with the classic approach, with our Multi-View Gaze Block (MVGB) efficiently extracting global, channel and high-frequency  local information. 

Given \( dense\_channel \) is the number of channels in the dense layers and \(depth\) is the number of layers in the dense block.

For each layer \( i \) in the Dense-TS:
\begin{itemize}
    \item The number of input channels to the \(i\)-th Two-Stage layer is \( C_{i} = dense\_channel \times (i + 1) \).
    \item The number of output channels from the \(i\)-th adjust convolutional layer is \( C_{out} = dense\_channel \).
\end{itemize}

The output of each TS-MVGB is a serial connection of time-MVGB and freq-MVGB:
\vspace{-0.1cm}
\begin{equation}
\text{TS-MVGB}_i(x) = \text{freq-MVGB} \left( \text{time-MVGB} \left( x \right) \right)
\end{equation}

Given the input feature map \( x \):
\vspace{-0.1cm}
\begin{equation}
\text{skip}_0 = x
\end{equation}

For \( i = 0, 1, \ldots, \text{depth} - 1 \):
\vspace{-0.1cm}
\begin{equation}
\text{skip}_i = \text{Concat} \left( \text{Adjust-Conv}(\text{TS-MVGB}_i(\text{skip}_{i-1})), \text{skip}_{i-1} \right)
\end{equation}

The final output of the Dense-TS is:
\vspace{-0.1cm}
\begin{equation}
\text{output} = 0.2 * \text{skip}_{\text{depth}-1} + x
\end{equation}

Additionally, after each Two-Stage module, a convolutional layer as \text{Adjust-Conv} is used to adjust the channel count and further integrate time-frequency information. Beyond dense connections, we add an extra residual connection. Finally, features undergo L-Sigmoid \cite{mpsenet_interspeech} activation and 1×1 convolution to decode the magnitude spectrum mask, which is then multiplied by the noisy magnitude spectrum to produce the network's output.

\subsection{Multi-View Gaze Block}
Inspired by NAFnet \cite{NAFnetchen2022simple} and Conformer \cite{conformer}, we designed the Multi-View Gaze Block (MVGB). As shown in Fig. \ref{fig:architecture} (d), the MVGB consists of three sub-modules: Large Kernel Extractor (LKE), Learnable-Sigmoid Gate (LSG), Channel Attention (CA).

\begin{figure}[t]
  \centering
  \includegraphics[width=\linewidth]{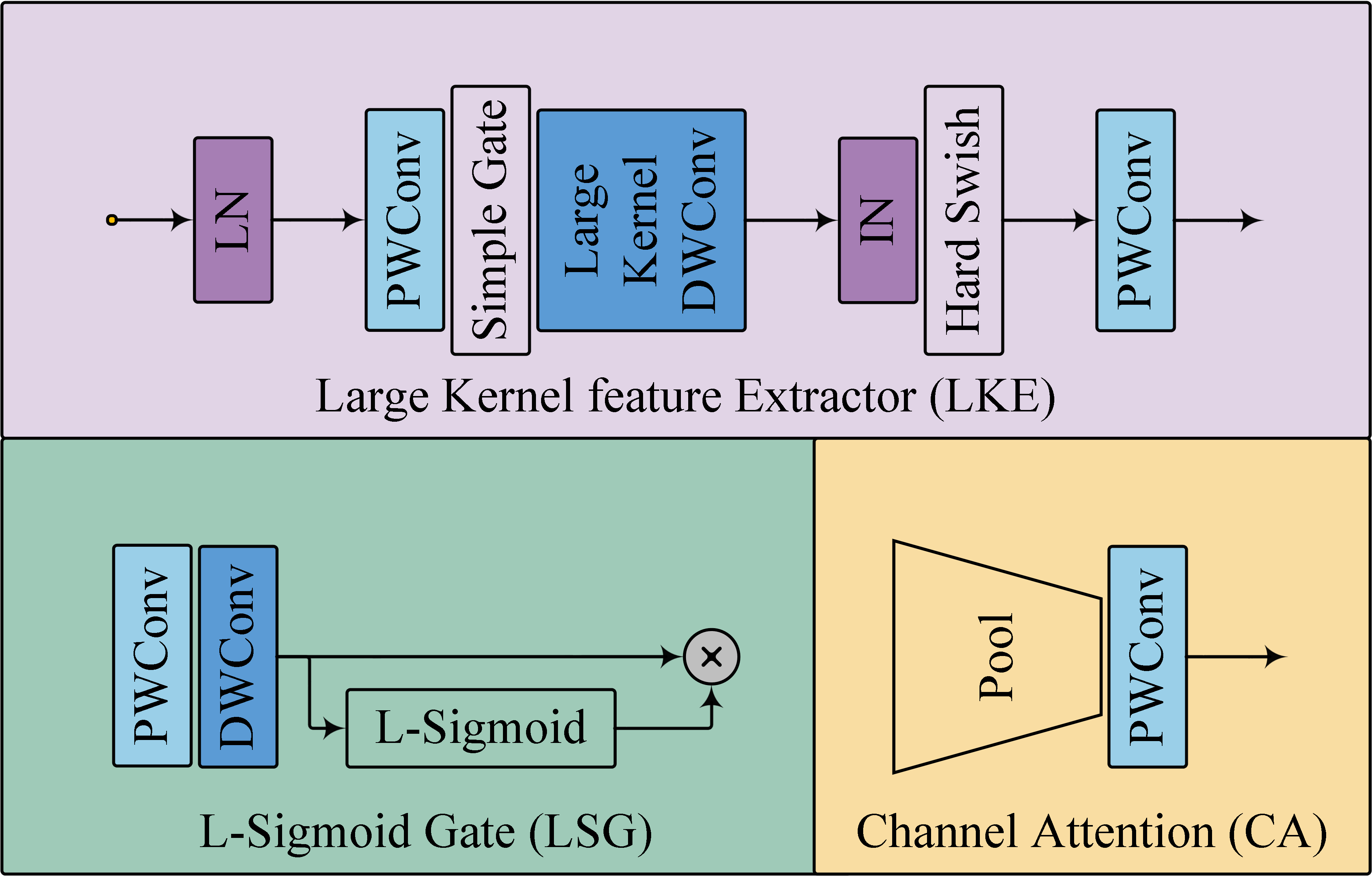}
  \caption{The internal details of the LKE, LSG and CA.}
  \label{fig:LKELSGCA}
  \vspace{-0.4cm}
\end{figure}
\textbf{Large Kernel Extractor (LKE)}
The Large Kernel Extractor (LKE) is responsible for establishing a global perspective, drawing inspiration from the CNN component of Conformer. As shown in Fig. \ref{fig:LKELSGCA}, LKE is composed of a sequence of pointwise convolutions, Simple Gate \cite{NAFnetchen2022simple}, large kernel depthwise convolutions \cite{mobilenets,conformer}, Instance Norm, HardSwish \cite{Hard-Swish}, and pointwise convolutions. The large kernel depthwise convolutions provide a large receptive field to capture long-range low-frequency features, while Simple Gate and HardSwish are designed with lightweight considerations.

The input \( x \) is normalized and then divided into two branches. 

\textbf{Channel Attention (CA)}
The first branch is Channel Attention (CA) \cite{NAFnetchen2022simple}, which reallocates weights to each channel, allowing the module to focus more on important channels while suppressing less significant ones, resulting in a more precise feature representation \cite{channelattensqueeze}.

\textbf{Learnable-Sigmoid Gate (LSG)}
The LSG shown in Fig. \ref{fig:LKELSGCA} is responsible for establishing local spatial perspectives, consists of depth-wise convolution, point-wise convolution, and Learnable-Sigmoid activation function. This branch is inspired by the Learnable-Sigmoid activation function used in MP-SENet\cite{mpsenet_interspeech} for generating masks. Its flexibility makes it crucial for generating accurate masks. This branch focuses on modeling local information within the feature map, enhancing high-frequency details and spatial invariance in the spectrum. 

Finally, by multiplying the channel weights and spatial weights obtained from the two branches, Multi-view Gazed features are generated, which are then further fused through pointwise convolution.


\subsection{Loss Functions}

\begin{figure}[t]
  \centering
  \includegraphics[width=\linewidth]{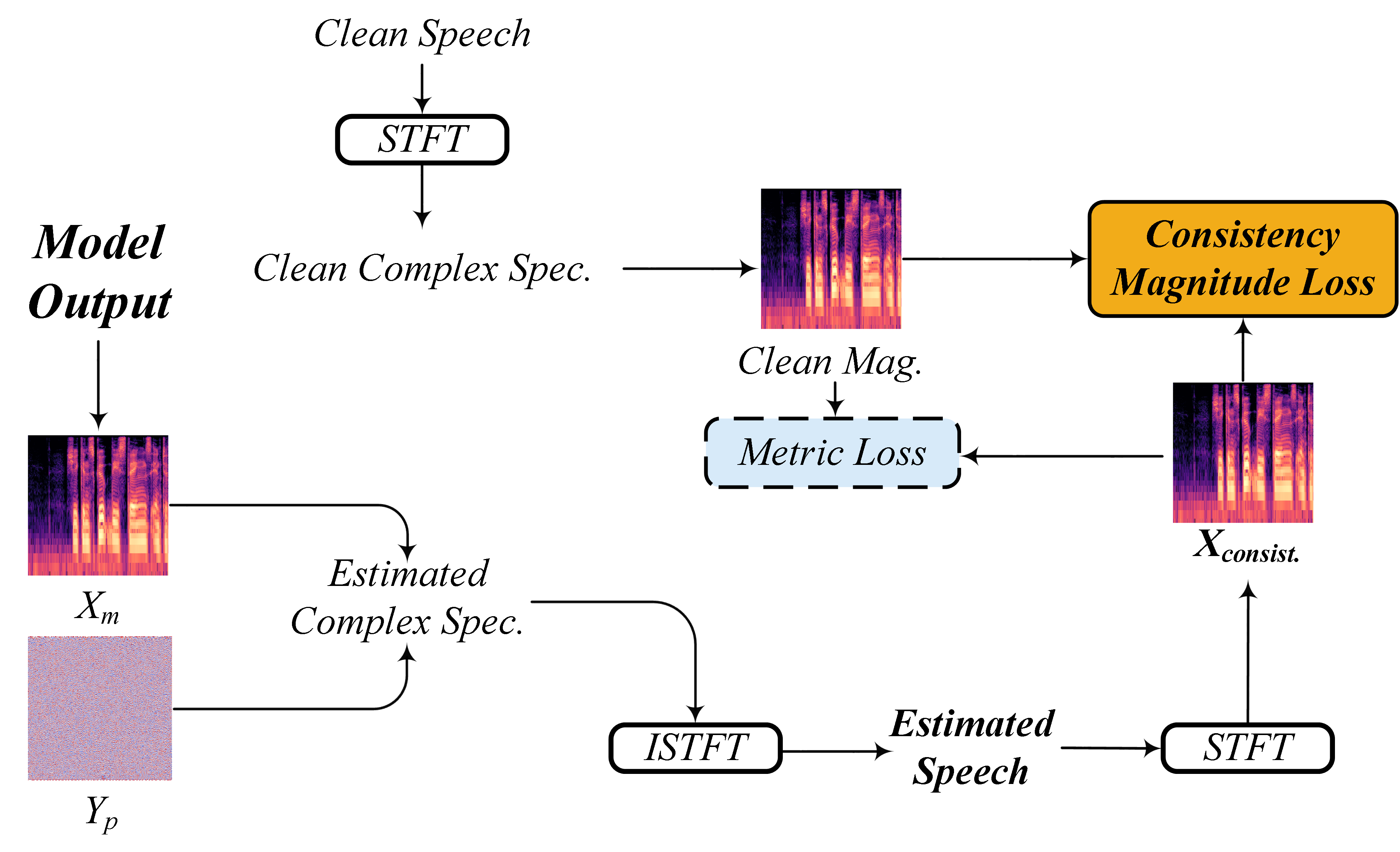}
  \caption{Training criteria of the proposed Dense-TSNet.}
  \label{fig:loss}
  \vspace{-0.4cm}
\end{figure}

Considering that lightweight networks cannot fully fit complex multi-domain features due to having fewer parameters compared to state-of-the-art (SOTA) methods \cite{DB-AIATinterspeech12,cmgan_interspeech,se-mamba,mpsenet_interspeech,wang23DPCFCS_interspeech,TridentSE_interspeech}, we chose not to use methods based on complex spectra or phase spectra. Instead, we designed a loss function that focuses on aspects more crucial to perceptual quality, aiming to minimize the distance between the estimated magnitude spectrum and the clean magnitude spectrum. Inspired by SCP-GAN \cite{scp}, since the Short-Time Fourier Transform (STFT) space differs from the space of the actual speech after the Inverse STFT (ISTFT). Previous methods based on the complex domain achieve consistency loss by computing the minimum distance between the complex spectra of speech after ISTFT and STFT, and the complex spectra output by the network. We directly compute the minimum distance between the magnitude spectrum of the twice-STFTed speech and the clean speech magnitude spectrum, as shown in Fig. \ref{fig:loss}. This loss $L _{Mag_{Consis.}}$is defined as Consistency Magnitude Loss:

\vspace{-0.2cm}
\begin{equation}
L _{Mag_{Consis.}}= \mathbb{E}_{X _m, X_{consis.}}\left[\left\| X _{ m } - X _{consis.}\right\|_2^2\right].
\end{equation}

Additionally, we validated the impact of using Metric Loss, which integrates objective evaluation metrics as loss functions, thereby improving model performance on perceptual quality scores. The Metric Loss is computed using the PESQ metric \cite{cmgan_interspeech,metricgan,metricgan+} via a discriminator:

\vspace{-0.2cm}
\begin{equation}
\begin{aligned}
L _{D} &= \mathbb{E}_{X _{m}}\left[\left\|D\left(X _{m}, X _{m}\right) - 1\right\|_2^2\right] \\
&\quad + \mathbb{E}_{X _{m}, X_{consis.}}\left[\left\|D\left(X _{m}, X _{consis.}\right) - Q_{PESQ}\right\|_2^2\right].
\end{aligned}
\end{equation}
\vspace{-0.2cm}
\begin{equation}
L _{\text{Metric}} = \mathbb{E}_{X _{m}, X_{consis.}}\left[\left\|D\left(X _{m}, X _{consis.}\right) - 1\right\|_2^2\right].
\end{equation}

Finally, we assign weights $\lambda_1$ and $\lambda_2$ to the two losses as Generator Loss, as $\lambda_2$ is set to $0$ by default:
\vspace{-0.1cm}
\begin{equation}
L _{G} = \lambda_1 L _{Mag_{Consis.}} + \lambda_2 L _{\text{Metric}}.
\end{equation}

\section{Experiments}

\subsection{Datasets and Model Setup}
For our experiments, we utilized the VoiceBank+DEMAND dataset \cite{voicebank}, known for its high-fidelity utterances. Speech samples were segmented into two-second intervals for the short-time Fourier transform, employing an FFT size of 400, a window length of 400, and a hop length of 100. The training process was conducted with a batch size of 2, using the AdamW optimizer \cite{kingma2014adam}, and was carried out for up to 1,000,000 steps.

\subsection{Evaluation Metrics}
To evaluate the quality of the denoised speech, we employed widely recognized metrics: PESQ \cite{pesq}, which ranges from -0.5 to 4.5, MOS-based metrics (CSIG, CBAK, COVL) with a range of 1 to 5. Higher scores on these metrics reflect superior performance.

\subsection{Result}
\vspace{-0.4cm}
\begin{figure}[ht]
    \centering
    \begin{subfigure}[b]{0.23\textwidth}
        \centering
        \includegraphics[width=\textwidth]{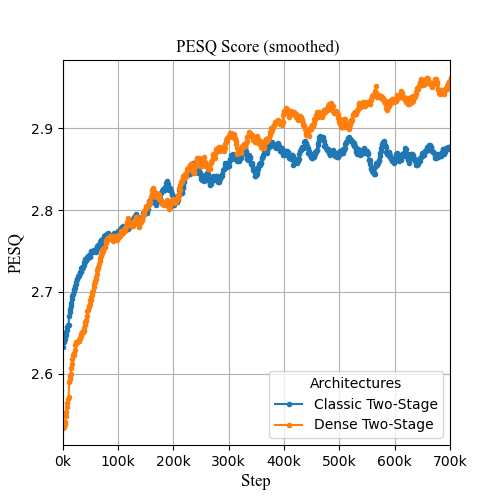}
        \caption{PESQ Score}
        \label{fig:image1}
    \end{subfigure}\hfill
    \begin{subfigure}[b]{0.23\textwidth}
        \centering
        \includegraphics[width=\textwidth]{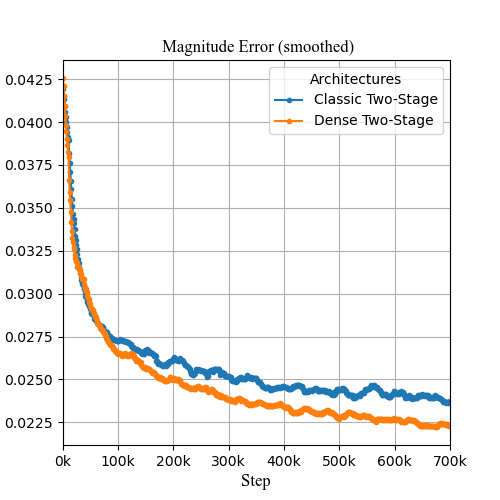}
        \caption{Magnitude Error}
        \label{fig:image2}
    \end{subfigure}\hfill
    \caption{Training curves of Dense-TS and Classic-TS}
    \label{fig:Training_curves}
\end{figure}
\vspace{-0.4cm}
Fig. \ref{fig:Training_curves} shows the training curves for the 6-channel version of the classic Two-Stage framework and the 4-channel version of the Dense-TS framework (with similar parameter counts), both of which use the MVGB approach we proposed. According to the PESQ curves, while the classic Two-Stage structure converges faster in the early stages, the Dense-TS framework can achieve a better performance in the later stages. This improvement is attributed to the smoother loss landscape provided by the dense connections in the narrower model \cite{loss_landscape}.

    

    
\vspace{-0.1cm}
\begin{table}[htbp]
    \centering
    \caption{Comparison with other methods on VoiceBank+DEMAND dataset. ``-'' denotes the result is not provided in the original paper.}
    \label{tab:compare}
    \setlength{\tabcolsep}{3pt} 
    \begin{tabular}{lcccccc}
    \hline
    Model & PESQ & CSIG & CBAK & COVL & Para. & MACs \\
    \hline
    TSTNN \cite{tstnn} & 2.96 & 4.33 & 3.53 & 3.67 & 0.92M  & - \\
    PHASEN \cite{phasen}  & 2.99 & 4.18 & 3.45 & 3.50 & 20.9M & - \\
    MANNER-S-5.3GF \cite{mannerlit} & 3.06 & 4.42 & 3.58 & 3.77 & 0.90M & - \\
    CCFNet(Lite) \cite{ccfnet} & 2.87 & - & - & - & 160K & 390M \\
    DeepFilterNet \cite{deepfilternet} & 2.81 & 4.14 & 3.31 & 3.46 & 1.78M & \textbf{350M} \\
    \hline
    Classic TS & 2.96 & 4.45 & 3.58 & 3.78 & \textbf{14K} & 388M \\
    \textbf{Dense-TSNet} & \textbf{3.05} & \textbf{4.51} & \textbf{3.58} & \textbf{3.86} & \textbf{14K} & 356M \\ 
    \hline
    \end{tabular}
\end{table}
\vspace{-0.1cm}

Table \ref{tab:compare} shows the performance comparison of our method with the classic Two-Stage structure and some lightweight algorithms. Our method achieves competitive performance with extremely low parameters.
\vspace{-0.1cm}
\begin{table}[htbp]
    \centering
    \caption{Ablation Study Results on the VoiceBank + DEMAND Dataset}
    \label{tab:ablation}
    \setlength{\tabcolsep}{6pt} 
    \begin{tabular}{lccccc}
    \hline
    Model & PESQ & CSIG & CBAK & COVL & SSNR \\
    \hline
    \textbf{Dense-TSNet} & \textbf{3.05} & \textbf{4.51} & \textbf{3.58} & \textbf{3.86}  & 8.16 \\
\hline
w/o LKE & 2.79 & 4.31 & 3.47 & 3.61 & 8.62 \\
w/o CA & 2.86 & 4.38 & 3.53 & 3.69 & 8.91 \\
w/o LSG & 2.97 & 4.46 & 3.57 & 3.79 & 8.69 \\
w/o Consistency Loss & 2.96 & 4.45 & 3.58 & 3.79 & \textbf{8.79} \\
    \hline
    \end{tabular}
    
\end{table}\
\vspace{-0.1cm}

Table \ref{tab:ablation} presents the ablation experiments for the MVGB module design, illustrating the effects of LKE (global view) as well as CA (channel view) and LSG (local view). Additionally, we validated the effects of the Consistency method on magnitude loss. 

\vspace{-0.1cm}
\begin{table}[htbp]
    \centering
    \caption{Study of Loss Function.}
    \label{tab:loss}
    \setlength{\tabcolsep}{6pt} 
    \begin{tabular}{lcccc}
    \hline
    Loss & PESQ & error mag. & error pha. & error com.\\
    \hline
    Only Consistency Mag. & 3.05 & \textbf{0.022} & \textbf{2.56} & 0.13 \\
    P = 1 & 3.28 & 0.057 & 2.73 & 0.15 \\
    P = 200 & \textbf{3.41} & 0.163 & 2.95 & 0.21 \\
    P = 400 & 3.04 & 0.800 & 3.95 & 0.41 \\
    Only complex  & 2.60 & 0.075 & 2.73 & \textbf{0.10} \\
    \hline
    \end{tabular}
\end{table}
\vspace{-0.1cm}

\begin{figure}[ht]
    \centering
    \begin{subfigure}[b]{0.23\textwidth}
        \centering
        \includegraphics[width=\textwidth]{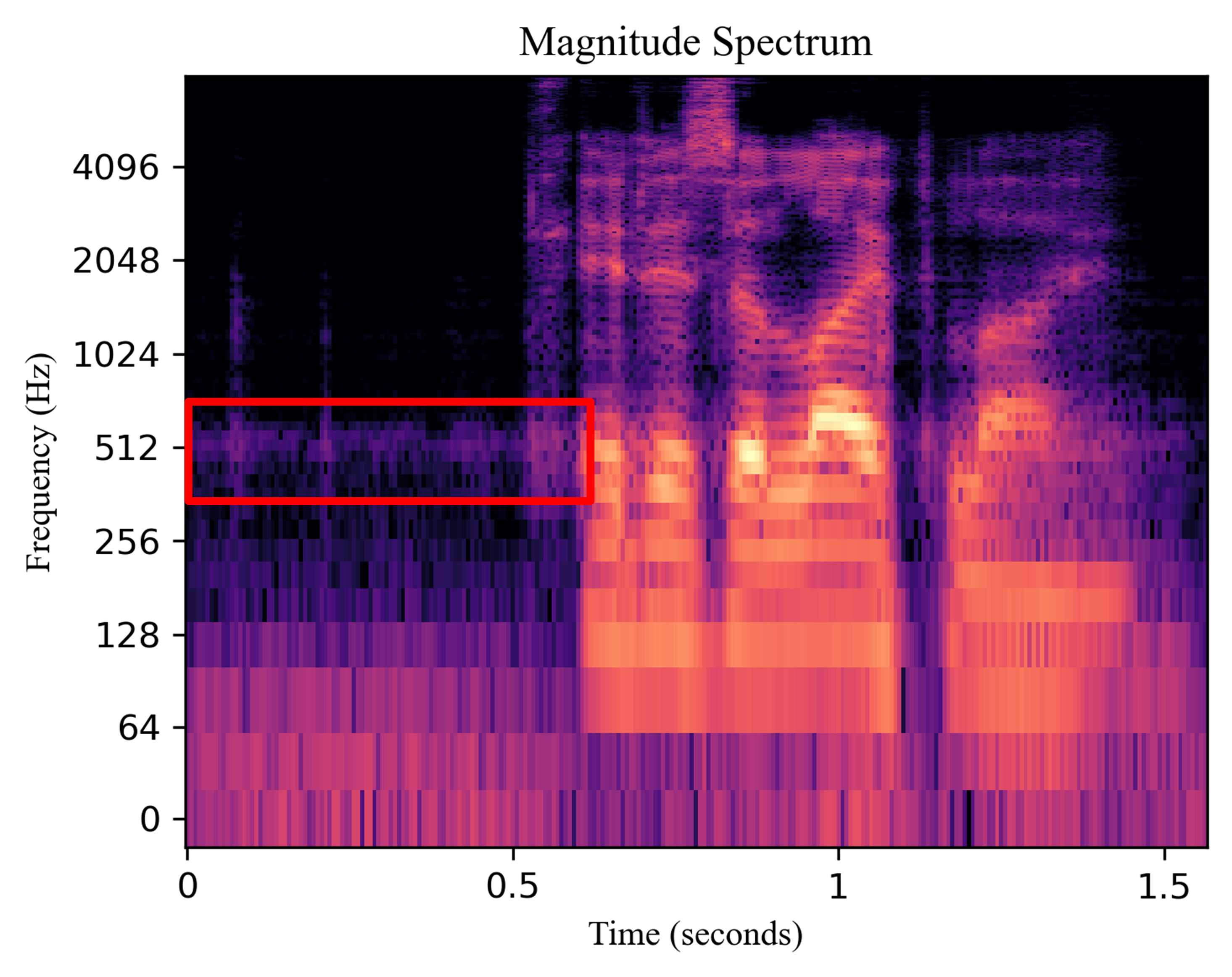}
        \caption{P=200}
        \label{fig:image1}
    \end{subfigure}\hfill
    \begin{subfigure}[b]{0.23\textwidth}
        \centering
        \includegraphics[width=\textwidth]{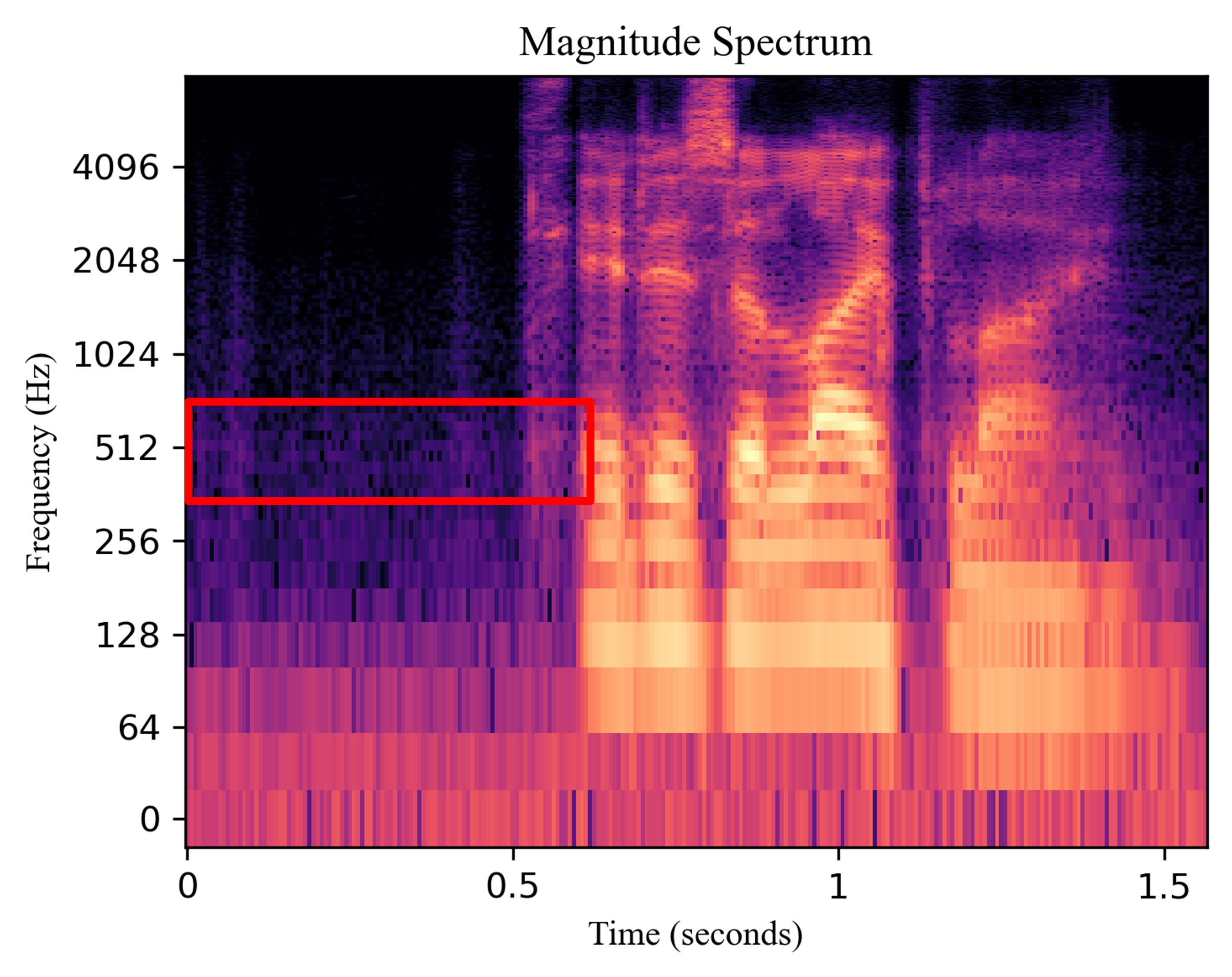}
        \caption{$L _{Mag_{Consis.}}$ only}
        \label{fig:image2}
    \end{subfigure}\hfill
    \caption{Training curves of Dense-TS and Classic-TS}
    \label{fig:mag._spec.}
\end{figure}

In Table \ref{tab:loss}, to validate the effect of $L_{\text{Metric}}$, we defined $P = \frac{\lambda_2}{\lambda_1}$, where $\lambda_1$ is the loss weight for $L_{\text{Mag}_{\text{Consis.}}}$, and $\lambda_2$ is the loss weight for $L_{\text{Metric}}$. We compared the results with those obtained using only the complex loss. Although the PESQ score increases with $P$, the distances between amplitude, phase, and the complex spectrum also increase. Moreover, when $P$ is too large (e.g., $P = 400$), training fails. For instance, as shown in Fig. \ref{fig:mag._spec.}, a pseudohallucination distortion is introduced at $P = 200$, affecting the perceptual quality. Given these observations, we do not employ $L_{\text{Metric}}$ in Dense-TS.

\vspace{-0.4cm}
\vspace{-0.4cm}

\section{Conclusions}

In this work, we have introduced Dense-TSNet, an ultra-lightweight speech enhancement network. The core of our approach is the Dense-Two-Stage (Dense-TS) architecture and the Multi-View Gaze Block (MVGB). Furthermore, we have explored the influence of the loss function. Dense-TSNet demonstrates exceptional efficacy with a compact model size of approximately 14K parameters, representing a substantial advancement for applications in resource-constrained environments.In the future, we will continue to explore methods for further compressing model size.

\bibliographystyle{ieeetr} 
\bibliography{mybib}

\begin{thebibliography}{10}

\bibitem{pascual2017segan}
S.~Pascual, A.~Bonafonte, and J.~Serr{\`a}, ``Segan: Speech enhancement generative adversarial network,'' {\em Interspeech 2017}, 2017.

\bibitem{tstnn}
K.~Wang, B.~He, and W.-P. Zhu, ``Tstnn: Two-stage transformer based neural network for speech enhancement in the time domain,'' in {\em ICASSP 2021-2021 IEEE International Conference on Acoustics, Speech and Signal Processing (ICASSP)}, pp.~7098--7102, IEEE, 2021.

\bibitem{dilateddensenet}
A.~Pandey and D.~L. Wang, ``Densely connected neural network with dilated convolutions for real-time speech enhancement in the time domain,'' in {\em IEEE International Conference on Acoustics, Speech and Signal Processing (ICASSP)}, ({Barcelona, Spain}), pp.~6629--6633, 2020.

\bibitem{cmgan_interspeech}
R.~Cao, S.~Abdulatif, and B.~Yang, ``{CMGAN: Conformer-based Metric GAN for Speech Enhancement},'' in {\em Proc. Interspeech 2022}, pp.~936--940, 2022.

\bibitem{mpsenet_interspeech}
Y.-X. Lu, Y.~Ai, and Z.-H. Ling, ``{MP-SENet: A Speech Enhancement Model with Parallel Denoising of Magnitude and Phase Spectra},'' in {\em Proc. INTERSPEECH 2023}, pp.~3834--3838, 2023.

\bibitem{metricgan}
S.-W. Fu, C.-F. Liao, Y.~Tsao, and S.-D. Lin, ``Metricgan: Generative adversarial networks based black-box metric scores optimization for speech enhancement,'' in {\em International Conference on Machine Learning}, pp.~2031--2041, PMLR, 2019.

\bibitem{metricgan+}
S.-W. Fu, C.~Yu, T.-A. Hsieh, P.~Plantinga, M.~Ravanelli, X.~Lu, and Y.~Tsao, ``Metricgan+: An improved version of metricgan for speech enhancement,'' {\em arXiv e-prints}, pp.~arXiv--2104, 2021.

\bibitem{vaswani2017attention}
A.~Vaswani, N.~Shazeer, N.~Parmar, J.~Uszkoreit, L.~Jones, A.~N. Gomez, {\L}.~Kaiser, and I.~Polosukhin, ``Attention is all you need,'' {\em Advances in neural information processing systems}, vol.~30, 2017.

\bibitem{mamba}
A.~Gu and T.~Dao, ``Mamba: Linear-time sequence modeling with selective state spaces,'' {\em arXiv preprint arXiv:2312.00752}, 2023.

\bibitem{se-mamba}
R.~Chao, W.-H. Cheng, M.~La~Quatra, S.~M. Siniscalchi, C.-H.~H. Yang, S.-W. Fu, and Y.~Tsao, ``An investigation of incorporating mamba for speech enhancement,'' {\em arXiv preprint arXiv:2405.06573}, 2024.

\bibitem{loss_landscape}
H.~Li, Z.~Xu, G.~Taylor, C.~Studer, and T.~Goldstein, ``Visualizing the loss landscape of neural nets,'' {\em Advances in neural information processing systems}, vol.~31, 2018.

\bibitem{CNNgradient}
Y.~LeCun, L.~Bottou, Y.~Bengio, and P.~Haffner, ``Gradient-based learning applied to document recognition,'' {\em Proceedings of the IEEE}, vol.~86, no.~11, pp.~2278--2324, 1998.

\bibitem{wang23DPCFCS_interspeech}
J.~Wang, ``{Efficient Encoder-Decoder and Dual-Path Conformer for Comprehensive Feature Learning in Speech Enhancement},'' in {\em Proc. INTERSPEECH 2023}, pp.~2853--2857, 2023.

\bibitem{DB-AIATinterspeech12}
G.~Yu {\em et~al.}, ``Dual-branch attention-in-attention transformer for single-channel speech enhancement,'' in {\em IEEE International Conference on Acoustics, Speech and Signal Processing (ICASSP)}, ({Singapore}), pp.~7847--7851, 2022.

\bibitem{NAFnetchen2022simple}
L.~Chen, X.~Chu, X.~Zhang, and J.~Sun, ``Simple baselines for image restoration,'' in {\em European Conference on Computer Vision}, pp.~17--33, Springer, 2022.

\bibitem{conformer}
A.~Gulati {\em et~al.}, ``Conformer: Convolution-augmented transformer for speech recognition,'' in {\em Proc. {INTERSPEECH} 2020 -- 21\textsuperscript{st} Annual Conference of the International Speech Communication Association}, ({Shanghai, China}), pp.~5036--5040, 2020.

\bibitem{mobilenets}
A.~G. Howard, ``Mobilenets: Efficient convolutional neural networks for mobile vision applications,'' {\em arXiv preprint arXiv:1704.04861}, 2017.

\bibitem{Hard-Swish}
R.~Avenash and P.~Viswanath, ``Semantic segmentation of satellite images using a modified cnn with hard-swish activation function.,'' in {\em VISIGRAPP (4: VISAPP)}, pp.~413--420, 2019.

\bibitem{channelattensqueeze}
J.~Hu, L.~Shen, and G.~Sun, ``Squeeze-and-excitation networks,'' in {\em Proceedings of the IEEE conference on computer vision and pattern recognition}, pp.~7132--7141, 2018.

\bibitem{TridentSE_interspeech}
D.~Yin, Z.~Zhao, C.~Tang, Z.~Xiong, and C.~Luo, ``{TridentSE: Guiding Speech Enhancement with 32 Global Tokens},'' in {\em Proc. INTERSPEECH 2023}, pp.~3839--3843, 2023.

\bibitem{scp}
V.~Zadorozhnyy, Q.~Ye, and K.~Koishida, ``Scp-gan: Self-correcting discriminator optimization for training consistency preserving metric gan on speech enhancement tasks,'' {\em arXiv preprint arXiv:2210.14474}, 2022.

\bibitem{voicebank}
C.~Valentini-Botinhao, X.~Wang, S.~Takaki, and J.~Yamagishi, ``Investigating rnn-based speech enhancement methods for noise-robust text-to-speech.,'' in {\em SSW}, pp.~146--152, 2016.

\bibitem{kingma2014adam}
D.~P. Kingma and J.~Ba, ``Adam: A method for stochastic optimization,'' {\em arXiv preprint arXiv:1412.6980}, 2014.

\bibitem{pesq}
A.~W. Rix, J.~G. Beerends, M.~P. Hollier, and A.~P. Hekstra, ``Perceptual evaluation of speech quality (pesq)-a new method for speech quality assessment of telephone networks and codecs,'' in {\em 2001 IEEE international conference on acoustics, speech, and signal processing. Proceedings (Cat. No. 01CH37221)}, vol.~2, pp.~749--752, IEEE, 2001.

\bibitem{phasen}
D.~Yin, C.~Luo, Z.~Xiong, and W.~Zeng, ``Phasen: A phase-and-harmonics-aware speech enhancement network,'' in {\em Proceedings of the AAAI Conference on Artificial Intelligence}, vol.~34, pp.~9458--9465, 2020.

\bibitem{mannerlit}
W.~Shin, H.~J. Park, J.~S. Kim, B.~H. Lee, and S.~W. Han, ``Multi-view attention transfer for efficient speech enhancement,'' {\em arXiv preprint arXiv:2208.10367}, 2022.

\bibitem{ccfnet}
F.~Dang, H.~Chen, Q.~Hu, P.~Zhang, and Y.~Yan, ``First coarse, fine afterward: A lightweight two-stage complex approach for monaural speech enhancement,'' {\em Speech Communication}, vol.~146, pp.~32--44, 2023.

\bibitem{deepfilternet}
H.~Schroter, A.~N. Escalante-B, T.~Rosenkranz, and A.~Maier, ``{DeepFilterNet: A low complexity speech enhancement framework for full-band audio based on deep filtering},'' in {\em ICASSP 2022-2022 IEEE International Conference on Acoustics, Speech and Signal Processing (ICASSP)}, pp.~7407--7411, IEEE, 2022.

\end{thebibliography}



\end{document}